\documentclass{iopart}
\usepackage{iopams}
\usepackage[utf8]{inputenc}
\usepackage{amsfonts}
\usepackage{bbold}
\usepackage{amsmath}
\usepackage{amssymb}
\usepackage[a4paper,left=23mm,top=20mm,right=23mm,bottom=15mm]{geometry}
\usepackage{graphics}
\usepackage{graphicx} 
\usepackage[export]{adjustbox}

\usepackage{hyperref}
\usepackage{listings}
\usepackage{booktabs}
\usepackage{xcolor}
\usepackage{comment}
\usepackage{dsfont}

\begin{document} 

\title{Continuous Variable Structured Collision Models}

\author{Anton Corr}
\address{%
 Centre for Quantum Materials and Technology, Queen’s University Belfast, BT7 1NN, United Kingdom 
}%
\address{%
\textit{Author to whom any correspondance should be addressed}
}%

\author{Stefano Cusumano}
 \address{Dipartimento di Fisica ‘Ettore Pancini’, Università degli Studi di Napoli Federico II, Via Cintia 80126, Napoli, Italy
 }%
\address{
 INFN, Sezione di Napoli, Italy
}

\author{Gabriele De Chiara}%
 
\address{%
 Centre for Quantum Materials and Technology, Queen’s University Belfast, BT7 1NN, United Kingdom 
}%
\address{Física Teòrica: Informació i Fenòmens Quàntics, Departament de Física, Universitat Autònoma de Barcelona, 08193 Bellaterra, Spain}

\author{Email:}%

\address{%
\href{mailto:acorr09@qub.ac.uk}{acorr09@qub.ac.uk}
}%

\begin{abstract}
Quantum collision models allow for the dynamics of open quantum systems to be described by breaking the environment into small segments, typically consisting of non-interacting harmonic oscillators or two-level systems. This work introduces structure within these environmental units via spring-like interactions between $N$ coupled oscillators in a ring structure, initially prepared in a thermal state. Two models of interest are examined. The first highlights a case in which a continuous time evolution can be obtained, wherein the system interacts with the environment via a beam-splitter-like, energy-preserving, interaction. The resulting dynamics are analogous to those due to interactions with unstructured units prepared as squeezed thermal states. The second model highlights a case in which the continuous time limit for the evolution cannot be taken generally, requiring instead discrete-time propagation. Special cases in which the continuous time limit can be taken are also investigated, alongside the addition of a secondary environment to induce a steady state. The first and second laws of thermodynamics are verified for both examples.
\end{abstract}

\maketitle

\section{Introduction}
The study of open quantum systems~\cite{10.1093/acprof:oso/9780199213900.001.0001, Bertolotti02012021}, i.e. systems that exchange energy, information and particles with their surrounding environments, plays a crucial role in the description of realistic quantum systems, as it is often impractical, or impossible, to neglect the effects of the environment. These environments typically contain a large number of degrees of freedom, making an exact analysis of such dynamics at best computationally difficult, and at worst impossible.

Quantum collision models \cite{CICCARELLO20221, e24091258, Cattaneo_2022, Campbell_2021} offer a method of simulating open dynamics by segmenting the large environment into smaller environmental units. These units then interact with the system sequentially for short timesteps, before being discarded post-interaction. These models allow for the dynamics of a large compilation of physical systems and environments to be examined while maintaining a small number of degrees of freedom.

Due to their versatility, collision models have become a popular method of investigating a wide variety of phenomena in both Markovian\cite{PhysRevLett.126.130403, Cattaneo_2022, CICCARELLO20221, PhysRevE.99.042103} and non-Markovian regimes \cite{PhysRevA.103.022202, Magalh_es_2023,_enya_a_2022, Ciccarello_2013},  quantum thermal machines \cite{DeChiaraPRR2020, Leitch_2022,Santiago-Garcia_2025, PhysRevLett.88.097905, PhysRevLett.127.100601, PhysRevA.102.042217, PiccionePRA2021, Hammam_2022, HammamPRR2024} and the behaviour of time crystals \cite{Campbell_2025, Carollo_2024}. Applications of such models can be found in the field of quantum optics \cite{qmetro-2017-0007,PhysRevResearch.2.043070, PhysRevA.95.053838, PhysRevA.97.053811},
quantum resources and information~\cite{PhysRevLett.115.120403, O’Connor_2024, PhysRevResearch.7.013145, PhysRevResearch.6.043321}, and many others \cite{PhysRevA.110.052421, PhysRevA.111.022209, PhysRevE.111.014115, campbell_global_2017,PhysRevA.111.032220}.

A key advantage of these models is the ability to study the thermodynamic properties of open systems~\cite{De_Chiara_2018, PhysRevA.91.022121, PhysRevA.101.062326, e23091198,  B_umer_2019, PhysRevA.100.042107,Rodrigues_2019, ShaghaghiPRE2022}, allowing for the investigation of the flow of the system's internal energy, alongside the corresponding thermodynamic heat and work flows. Furthermore, the entropy and entropy production are readily available~\cite{PhysRevA.98.032119,PhysRevX.7.021003}, allowing for the verification of the laws of thermodynamics.

In a typical collision model, the system and environmental units are considered to be composed of independent objects, such as harmonic oscillators or two-level systems. Recently, however, focus has been drawn to the idea of structured collision models, wherein interactions within the system or environment are included \cite{Cusumano_2024}. In this work, we present the case of a structured continuous-variable environment, studying the dynamics of a system of harmonic oscillators interacting with environmental units composed of an arbitrary number of interacting harmonic oscillators.

We examine two examples of interest in this work. In both, the environmental units consist of $N_E$ harmonic oscillators coupled via spring-like interactions in a ring structure (see Fig.~\ref{fig:Diagram}). The first example investigates the effects of coupling the system and environment through beamsplitter-like interactions. In this example, the continuous time limit for the evolution of the reduced state of the system can be taken, allowing for the corresponding Lyapunov equation for the covariance matrix to be derived. We find that preparing the structured environmental units in a thermal state leads to dynamics analogous to those produced through interactions with an unstructured environment initially prepared in a squeezed thermal state, with a modified effective temperature.

The second example considers spring-like interactions between the system and environment, which may result in divergences when taking the continuous time limit. We find that discrete time propagation must be used in the general case, with the continuous time limit only possible in special cases. We emphasise, however, that these cases are not achievable without the use of environmental units composed of a minimum of two oscillators. While this example does not result in a steady state, a method of inducing a steady state through the inclusion of a secondary unstructured environment interacting with the system via a beamsplitter-like interaction is also presented.

This work is organised such that the general model is derived in Sec.~\ref{sec; General Model}, including the structure of the collision model, the corresponding dynamics and the methods of deriving the relevant thermodynamic quantities. Sec.~\ref{sec; Examples} introduces the physical examples, with Sec.~\ref{sec; BS Example} presenting the beamsplitter-like system-environment interactions. Sec.~\ref{sec; Spring Example} then presents the spring-like system-environment interactions, including the general discrete time propagation and continuous time special cases. Outlooks on future work, alongside the conclusions drawn from the models, will be presented in Sec.~\ref{sec; Conclusions}. Some details on the analysis and notation used can be found in the appendices.

\section{General Model} \label{sec; General Model}
\begin{figure}
    \centering
    \includegraphics[width=\linewidth]{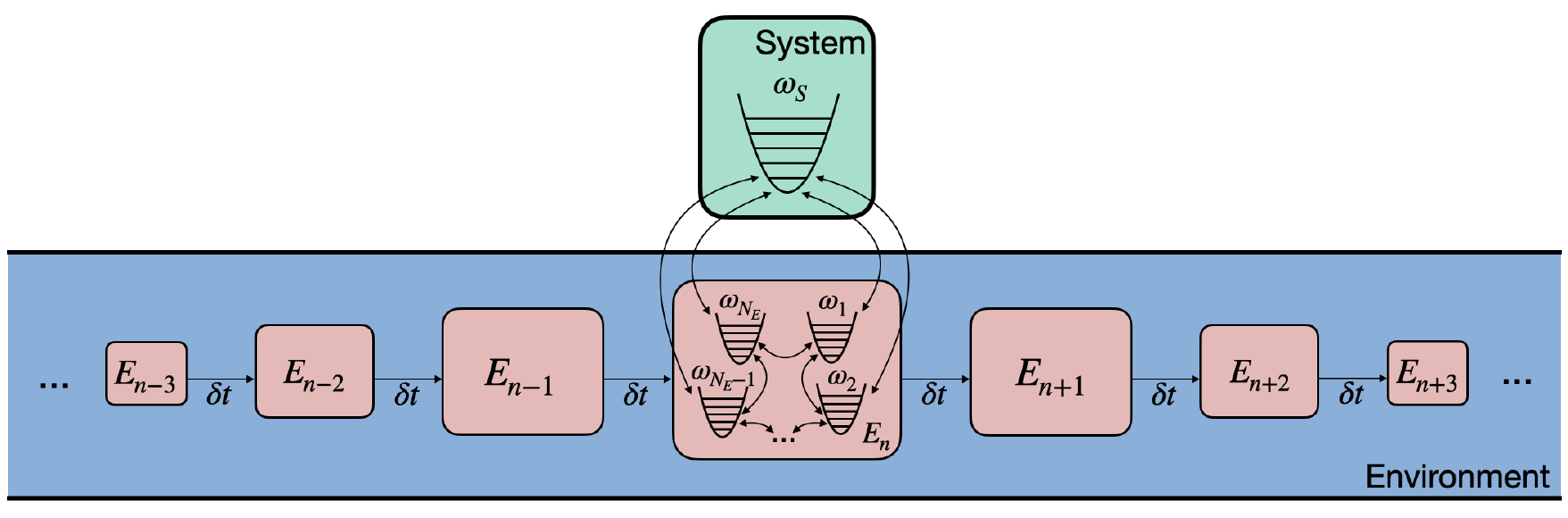}
    \caption{A system composed of a harmonic oscillator interacting with an environment composed of structured units, each containing $N_E$ interacting oscillators arranged in a ring, via a collision model. Each environmental unit interacts with the system for a timestep $\delta t$ before being replaced by an identical unit.}
    \label{fig:Diagram}
\end{figure}

Collision models, in general, allow for the study of the dynamics of a system $\mathcal{S}$ interacting with an environment $\mathcal{E}$. As the environment is typically large and complicated, tracking the interactions with the total environment may be conceptually difficult and computationally expensive. To combat this, collision models treat the environment as an infinitely large collection of, generally identically prepared, environmental units $E_i$. In the Markovian regime, the system interacts with each unit sequentially for a time $\delta t$, and the unit is discarded after the interaction to ensure the system's evolution is Markovian, see Fig.~\ref{fig:Diagram}. Often, it then becomes possible to derive a continuous Lyapunov equation for the system by taking the continuous limit of $\delta t \to 0$ after imposing some conditions on the interaction between the system and the environment. 

%
%

\subsection{System and Environment Structure} \label{sec; General Structure}
As a general case, we will consider a system of $N_S$ oscillators and an environment made of an infinite number of units, each composed of $N_E$ coupled oscillators. The system then interacts with each of these units sequentially for a time $\delta t$. The total Hamiltonian $H_{\text{tot}}$ of such a model is composed of the free Hamiltonian of the system oscillators, $H_S$, the Hamiltonian of the environmental unit, $H_E$, which includes the coupling terms present within the environment, and the system-environment interaction Hamiltonian, $H_{SE}$, such that
\begin{equation}
    H_{\text{tot}} = H_S + H_E + H_{SE}.
\end{equation}
The Hamiltonian of the system can be described in terms of the position and momentum operators of each oscillator $i$, namely $x_{S_i}$ and $p_{S_i}$ respectively, such that
\begin{equation}
    H_S = \frac{1}{2}\sum_{i=1}^{N_S} \left(\omega_{S_i}^2 x_{S_i}^2 + p_{S_i}^2\right). \label{eq:System Hamiltonian General}
\end{equation}
For simplicity, in this work we consider uncoupled system oscillators of unit mass $m$, with $\hbar = m = 1$. Furthermore, the position and momentum operators obey the standard commutation relation $[x_{S_i},p_{S_j}] = i\delta_{ij}$. Eq.~\eqref{eq:System Hamiltonian General} can be expressed in matrix notation as
\begin{equation}
  H_S = \frac{1}{2} R_S^\intercal M_S R_S,
\end{equation}
with $R_S$ denoting a column vector containing the position and momentum operators such that $R_S^\intercal = \begin{pmatrix}x_{S_1}, p_{S_1}, x_{S_2}, p_{S_2}, ..., x_{S_{N_S}}, p_{S_{N_S}}\end{pmatrix}$, and 
\begin{equation}
    M_S = 
    \bigoplus_{i=1}^{N_S} 
    \begin{pmatrix}
        \omega_{s_i}^2 & 0 \\
        0 & 1
    \end{pmatrix}.
\end{equation}

The Hamiltonian for each of the environmental units can be separated into two terms, namely the free Hamiltonian $H_E^{F}$ and the interaction Hamiltonian describing the internal interactions within the unit $H_E^I$, where the total $H_E$ is
\begin{equation}
    H_E = H_E^{F} + H_E^I. \label{eq:Environment Hamiltonian}
\end{equation}
The free Hamiltonian of each environmental unit can be written in a form analogous to Eq.~\eqref{eq:System Hamiltonian General}:
\begin{equation}
    H_E^F = \frac{1}{2}\sum_{i=1}^{N_E} \left(\omega_{E_i}^2 x_{E_i}^2 + p_{E_i}^2\right). \label{eq:H_E Free}
\end{equation}
with $x_{E_i}$ denoting the position operator of each environmental oscillator $i$, and $p_{E_i}$ the corresponding momentum operator obeying the analogous commutation relations, $[x_{E_i},p_{E_j}] = i\delta_{ij}$. The Hamiltonians $H_E^I$ and $H_{SE}$ are then dependent on the choice of internal environmental interactions and system-environment interactions, respectively. The environmental Hamiltonian can also be cast in matrix form, with
\begin{equation}
  H_E = \frac{1}{2} R_E^\intercal M_E R_E,
\end{equation}
where $R_E^\intercal = \begin{pmatrix}x_{E_1}, p_{E_1}, x_{E_2}, p_{E_2}, ..., x_{E_{N_E}}, p_{E_{N_E}}\end{pmatrix}$, and $M_E$ as the corresponding matrix of coefficients. 

In this work, we assume each environmental unit to be prepared initially in a thermal state of the form $\rho_E=\exp(-\beta_E H_E)/Z_E$ where $Z_E = \Tr[\exp(-\beta_E H_E)]$ and $\beta_E = 1/T_E$. For simplicity, units have been chosen such that $k_B = 1$, where $k_B$ is the Boltzmann constant. As only Gaussian states will be considered, it is convenient to introduce the covariance matrix, $\sigma$, defined through 
\begin{equation}
    \sigma_{ij} = \frac{1}{2}\langle R_iR_j + R_jR_i\rangle - \langle R_i \rangle \langle R_j \rangle. \label{eq:Covariance Definition}
\end{equation}
Here, $R_i$ refers to the system or environment's operators, while the notation $\langle R_i \rangle=\Tr[\rho R_i]$ denotes the average value of an operator, where $\rho$ is the relevant density matrix of the system, $\rho_S$, environmental unit $\rho_{E}$, or the combination of the system and environmental unit $\rho_{\rm{tot}}$. 

As $M_E$ is necessarily non-diagonal in the basis of $R_E$, due to the internal environmental interactions, the covariance matrix of the environment $\sigma_E$ in the thermal state is not  diagonal. It is possible, however, to define a normal-mode coordinate system $Q_E^\intercal = \begin{pmatrix} \tilde x_{E_1}, \tilde p_{E_1}, \tilde x_{E_2}, \tilde p_{E_2}, ..., \tilde x_{E_{N_E}}, \tilde p_{E_{N_E}}\end{pmatrix}$, through a transformation $G$ such that $Q_E = GR_E$, for which the covariance matrix is diagonal. The corresponding Hamiltonian can be expressed as
\begin{equation}
    H_E = \frac{1}{2}Q_E^\intercal \tilde{M}_E Q_E,
\end{equation}
where $\tilde{M}_E$ is a diagonal matrix defined by 
\begin{equation}
    \tilde{M}_E = G M_E G^\intercal. \label{eq:Matrix Form}
\end{equation}
This transformation matrix $G$ must be symplectic to ensure the commutation relations are preserved under the change of basis, i.e. $G^\intercal \Omega_{N_E} G = \Omega_{N_E}$, where
\begin{equation}
    \Omega_{N_E} = 
        \bigoplus_{i=1}^{N_E}
    \begin{pmatrix}
        0 & 1 \\
        -1 & 0
    \end{pmatrix}.
\end{equation}
In general, $\tilde{M}_E$ takes the form
\begin{equation}
    \tilde{M}_E =
    \bigoplus_{m=1}^{N_E}
    \begin{pmatrix}
        \tilde{\omega}_{E_m} & 0 \\
        0 & 1
    \end{pmatrix},
\end{equation}
where $\tilde{\omega}_{E_m}$ denotes the normal mode frequencies of the environmental unit.
For the interactions considered in this work, we find $G^\intercal G = \mathds{1}_{2N_E}$, with $\mathds{1}_{2N_E}$ as the identity matrix of dimension $2N_E$, where the columns of $G$ correspond to the orthogonal eigenvectors of $M_E$. One can then recover the basis of $R_E$ through
\begin{equation}
    \sigma_E = G^\intercal \tilde{\sigma}_E G.
\end{equation}

The total state of the system at time $t$ and an environmental unit can then be described through the covariance matrix $\sigma_{\text{tot}}(t)$,
\begin{equation}
    \sigma_{\text{tot}}(t) = \sigma_S(t) \oplus \sigma_E, \label{eq:sigma_tot definition}
\end{equation}
where $\sigma_S(t)$ is the covariance matrix corresponding to the system at a time $t$. In Eq.~\eqref{eq:sigma_tot definition}, we employed the direct sum as a consequence of assuming no initial correlations between the system and the unit.

For completeness, one can also recast the system-environment interaction Hamiltonian, $H_{SE}$ into its matrix form, producing
\begin{equation}
    H_{SE} = \frac{1}{2} R_{\text{tot}}^\intercal M_{SE} R_{\text{tot}},
\end{equation}
where $R_{\text{tot}} = R_S \oplus R_E$.

\subsection{System Dynamics} \label{sec; General Dynamics}
The Markovian collision model framework describes the system's interaction with an environment through a series of short interactions of duration $\delta t$ with the units of the environment. Each unit is discarded after the collision to ensure Markovianity. As in this work we consider the dynamics of Gaussian states, we are interested in the propagation of the system's first and second moments. The evolution of the second moments can be described through a symplectic operator $U$ such that 
\begin{equation}
    \sigma_{\text{tot}}(t + \delta t) = U\sigma_{\text{tot}}(t)U^\dagger \label{eq:sigma(t + dt)}
\end{equation}
where $\sigma_{\text{tot}}(t)$ is defined in Eq.~\eqref{eq:sigma_tot definition}. The time evolution operator $U$ is described as
\begin{equation}
    U = e^{\delta t D_{\text{tot}}},
\end{equation}
where the propagation matrix $D_{\text{tot}}$ satisfies
\begin{equation}
    \langle \dot{R}\rangle_{\text{tot}} = D_{\text{tot}}\langle R\rangle_{\text{tot}} \label{eq:R Dynamics}
\end{equation}
with $\langle \dot{R} \rangle_{\text{tot}} = \frac{d}{dt}\langle R \rangle _{\text{tot}}$. The elements of this propagation matrix $D_{\text{tot}}$ can be found directly using the Heisenberg equation for the system and environment position and momentum operators, or equivalently through $D_{\text{tot}} = \Omega_{N_{\text{tot}}} M_{\text{tot}}$, where 
\begin{equation}
    H_{\text{tot}} = \frac 12 R_{\text{tot}}^\intercal M_{\text{tot}}R_{\text{tot}}
\end{equation} 
and $N_{\text{tot}} = N_S + N_E$. 

Expanding Eq.~\eqref{eq:sigma(t + dt)} through the Taylor expansion to second order in $\delta t$ produces three superoperators:
\begin{equation}
    U\sigma_{\text{tot}}(t)U^\dagger \simeq \mathcal{U}_0[\sigma_{\text{tot}}(t)] + \mathcal{U}_1[\sigma_{\text{tot}}(t)] + \mathcal{U}_2[\sigma_{\text{tot}}(t)].
\end{equation}
The superoperator $\mathcal{U}_i[\sigma_{\text{tot}}(t)]$ describes the $i$-th order contributions from the Taylor expansion, with
\begin{eqnarray}
    \mathcal{U}_0[\sigma_{\text{tot}}(t)] &=& \sigma_{\text{tot}}(t),
    \\
    \mathcal{U}_1[\sigma_{\text{tot}}(t)] &=& \delta t\left[D_{\text{tot}}\sigma_{\text{tot}}(t) + \sigma_{\text{tot}}(t)D_{\text{tot}}^\intercal \right ],
    \\
    \mathcal{U}_2[\sigma_{\text{tot}}(t)] &=& \frac{\delta t^2}{2}  \left [2D_{\text{tot}}\sigma_{\text{tot}}(t)D_{\text{tot}}^\intercal + D_{\text{tot}}^2\sigma_{\text{tot}}(t) + \sigma_{\text{tot}}(t)D_{\text{tot}}^{\intercal 2} \right ].
\end{eqnarray}

As the environmental units are replaced between collisions, it is only necessary to retain the effect these superoperators have on the block of the covariance matrix that describes the system, i.e. $\sigma_S(t)$. This is mathematically equivalent to tracing out the degrees of freedom of the relevant environmental unit after each collision. By discarding the components outside of this block, these superoperators can be simplified to retain only the required matrices of dimensions $2N_S$. The dynamics of the system can then be described through $\sigma_S(t+\delta t) = \mathcal{L}\left [\sigma_{\text{tot}}(t)\right ]$ where $\mathcal{L}\simeq\mathcal{L}_0+\mathcal{L}_1+\mathcal{L}_2$
to second order in  $\delta t$, where

\begin{eqnarray}
    \mathcal{L}_0[\sigma_{S}(t)] &=& \sigma_S(t), \label{eq:Zero Order Superoperator}
    \\
    \mathcal{L}_1[\sigma_{S}(t)] &=& \delta t\left\{D_S \sigma_S(t) + \sigma_S(t)D^\intercal_S + [D_{SE}\sigma_{\text{tot}}(t)]_S + [\sigma_{\text{tot}}(t)D_{SE}^\intercal]_S\right\}, \label{eq:First Order Superoperator}
    \\
    \mathcal{L}_2[\sigma_S (t)] &=& \frac{\delta t^2}{2} \left\{\mathcal{L}^S_2[\sigma_S (t)] + \mathcal{L}^{SE}_2[\sigma_S (t)]+ \mathcal{L}^{S/SE}_2[\sigma_S (t)]\right\} \label{eq:Second Order Superoperator}.
\end{eqnarray}

The detailed derivation of these superoperators, along with the explicit expressions of $\mathcal{L}^S_2 , \mathcal{L}^{SE}_2$ and $\mathcal{L}^{S/SE}_2$, can be found in  \ref{App; System superoperators}.  We have introduced the notation $[A]_S$ to denote the partial trace of the environment through the discarding of elements of the covariance matrix outside of those corresponding to the system, retaining only matrix elements $A_{ij}$ for which $1\leq i \leq 2N_S$ and $1\leq j \leq 2N_S$. Here, $D_S$ is a matrix of dimension $2N_S$ describing the dynamics of the free Hamiltonian of the system such that $D_S = \Omega_{N_S} M_S$. $D_{SE}$ is then a matrix of dimension $2N_{\text{tot}}$, such that $D_{SE} = \Omega_{N_{\text{tot}}} M_{SE}$.

As we will see throughout the examples in Sec.~\ref{sec; Examples}, $D_{SE}$ is dependent on the choice of system-environment interaction Hamiltonian. The terms relating to the free evolution of the environment do not survive the tracing out of the environmental unit; instead the internal environmental couplings only affect the system through the initial state $\sigma_{\text{tot}}(t)$ and the $D_{SE}$ terms.

The discrete propagation of the system can then be described through
\begin{equation}
    \frac{\sigma_S(t + \delta t) - \sigma_{S}(t)}{\delta t} =  \frac{\mathcal{L}_1(\sigma_{\text{tot}}(t)) + \mathcal{L}_2(\sigma_{\text{tot}}(t))}{\delta t}. \label{eq: Pre continuous}
\end{equation}
The continuous time limit of this propagation is obtained by rescaling the coupling constants $g_{E_i}$ within $D_{SE}$ before taking the limit of $\delta t \to 0$, as shown in Sec.~\ref{sec; BS Example}, to form a Lyapunov equation. This is not always possible without divergences, however the following conditions are sufficient to ensure a finite Lyapunov equation:

\begin{enumerate}
    \item The second order term $\mathcal{L}_2$ must be a homogeneous function of the coupling constants $g_{E_i}$, where $g_{E_i}$ represents the coupling strength between the system and each oscillator $i$ within the environmental unit.\label{Condition i}
    \item The first order corrections to the system's free evolution appearing in $\mathcal{L}_1$ (see Eq.~\eqref{eq:First Order Superoperator}) must be zero: $[D_{SE}\sigma_{\text{tot}}(t)]_S = [\sigma_{\text{tot}}(t)D_{SE}^\intercal]_S = 0$. \label{Condition ii}
\end{enumerate}

In this work, we consider quadratic coupling Hamiltonians for which condition~\ref{Condition i} is always satisfied. In cases for which condition (\ref{Condition ii}) is satisfied, we find that the terms corresponding to the first order expansion in $\delta t$, $\mathcal{L}_1(\sigma_{\text{tot}}(t))$, describe the free evolution of the system, while the second order terms, $\mathcal{L}_2(\sigma_{\text{tot}}(t))$, result from the interaction of the system with the environment.

\subsection{Thermodynamic Quantities}\label{sec; Thermo Quantities}
The change in the system's internal energy $\Delta U(t)$, the heat $\Delta Q(t)$ exchanged with the environment and the associated thermodynamic work $\Delta W(t)$ are natural quantities of interest~\cite{Alicki_1979,PhysRevLett.88.097905,PhysRevX.7.021003,De_Chiara_2018, PhysRevE.99.042145,Cusumano_2024} that can be extracted from the dynamics described in Sec.~\ref{sec; General Dynamics}. These can be expressed through 
\begin{eqnarray}
    \Delta U(t) &=& \frac{1}{2} \Tr\left[\Delta \sigma_{\rm{tot}}(t) (M_S \oplus O_{2N_E}) \right],
\\
    \Delta Q(t) &=& -\frac{1}{2} \Tr\left[\Delta \sigma_{\rm{tot}}(t) (O_{2N_S} \oplus M_E) \right],
\\
    \Delta W(t) &=& \frac{1}{2} \Tr\left[\Delta \sigma_{\rm{tot}}(t) (M_S \oplus M_{E}) \right],
\end{eqnarray}
where $\Delta \sigma_{\rm{tot}}(t) = \sigma_{\rm{tot}}(t + \delta t) - \sigma_{\rm{tot}}(t)$ and $O_N$ denotes the zero matrix of dimension $N$. These quantities satisfy the first law of thermodynamics by construction, i.e.
\begin{equation}
    \label{eq:firstlaw}
    \Delta U(t) = \Delta Q(t) + \Delta W(t).
\end{equation}

One can also consider the von Neumann entropy of the system
\begin{equation}
    S_{\rm{VN}}(\rho_S) = -\Tr[\rho_S \ln\rho_S], \label{eq:Entropy w.r.t. density matrix}
\end{equation}
where $\rho_S$ is the reduced density operator of the system.
To remain consistent with the covariance matrix formalism described in Sec.~\ref{sec; General Structure}, it is instructive to cast Eq.~\eqref{eq:Entropy w.r.t. density matrix} in terms of the symplectic eigenvalues $\nu_k$ of  $\sigma_S(t)$, where $\nu_k$ are defined as the absolute value of the eigenvalues of $i\Omega_{N_S}\sigma_S(t)$. As the eigenvalues of $\sigma_S$ are paired such that each $\nu_k$ is 2-fold degenerate, we consider only the $N_S/2$ unique elements. This allows for the entropy to be calculated as \cite{alma991000777479708046}
\begin{equation}
    S_{\rm{VN}}(\rho_S(t)) = \sum_{k=1}^{N_S/2} S_e(\nu_k).
\end{equation}
$S_e(\nu_k)$ takes the form
\begin{equation}
    S_e(\nu_k) = \left(\nu_k + \frac{1}{2}\right)\ln\left(\nu_k + \frac{1}{2}\right) - \left(\nu_k - \frac{1}{2}\right)\ln\left(\nu_k - \frac{1}{2}\right).
\end{equation}

As information is lost between collisions due to the discarding and replacing of the environmental units, it is important to focus the analysis on the entropy production per collision, defined as
\begin{equation}
    \Sigma(t) = \Delta S_{\rm{VN}}(t) - \frac{\Delta Q(t)}{T_E}.
\end{equation}
The entropy production then allows for the verification of the second law of thermodynamics,
\begin{equation}
    \Sigma(t) \geq 0,
\end{equation}
 namely that the entropy production must be non-negative during time evolution.

\section{Examples} \label{sec; Examples}
In order to examine the effects of the interactions present within the environment, two key examples will be discussed in detail here. For each example, the results presented will consider a system consisting of a single harmonic oscillator of frequency $\omega_{S}$ (thus $N_S=1$), initially prepared in a thermal state of temperature $T_S$. The environmental units shall consist of $N_E$ oscillators, each of frequency $\omega_E$, coupled in a ring structure through spring-like interactions. Each unit as a whole is considered to be prepared in a thermal state at temperature $T_E$. The system-environment interactions will then be described as beamsplitter-like in Sec.~\ref{sec; BS Example}, and spring-like in Sec.~\ref{sec; Spring Example}.

\subsection{Initial States}
In both examples, we consider the same initial covariance matrix describing the system and environment $\sigma_{\text{tot}}(0)$. The system, described by the Hamiltonian defined in Eq.~\eqref{eq:System Hamiltonian General}, is initially prepared in a thermal state of the form $\rho_{S,{\rm th}}=\exp(-\beta_S H_S)/Z_S$, where $Z_S = \Tr[\exp(-\beta_S H_S)]$ and $\beta_S = 1/T_S$. The corresponding initial covariance matrix for the single oscillator system can then be written as
\begin{equation}
    \sigma_S(0) = \begin{pmatrix}
        \frac{1}{2 \omega_S}\coth(\frac{ \omega_S}{2T_S}) & 0 \\
        0 & \frac{\omega_S}{2} \coth(\frac{ \omega_S}{2T_S})
    \end{pmatrix},
\end{equation}
where the terms proportional to $\coth(\frac{ \omega_S}{2T_S})$ relate to the thermal population of the system.

We assume the environmental units to be arranged along a linear chain with nearest-neighbor interactions and periodic boundary conditions, i.e. a ring-like structure.  The corresponding Hamiltonian $H_E$ can be described by Eq.~\eqref{eq:Environment Hamiltonian} and Eq.~\eqref{eq:H_E Free}, where 
\begin{equation}
    H_E^I = \lambda_{I}\sum_{i=1}^{N_E} (x_{E_i} - x_{E_{i+1}})^2. \label{eq:H_EI}
\end{equation}
Here, $\lambda_I$ is the coupling constant controlling the strength of the interactions within the environment. We also take $x_{N_E +1} \equiv x_1$ to satisfy the periodicity of the boundary conditions. Following the steps outlined in Sec.~\ref{sec; General Structure}, this can be recast in matrix form akin to Eq.~\eqref{eq:Matrix Form}.

The covariance matrix $\sigma_E$ of the environmental unit before each collision can then be written as
\begin{equation}
    \sigma_E = G^\intercal \left(\bigoplus_{m=1}^{N_E} \tilde{\sigma}_{E_m}\right) G,
\end{equation}
where
\begin{equation}
        \tilde{\sigma}_{E_m} = 
        \begin{pmatrix}
        \frac{1}{2 \tilde{\omega}_{E_m}}\coth\left({\frac{\tilde{\omega}_{E_m}}{2T_E}}\right) & 0 \\
        0 & \frac{\tilde{\omega}_{E_m}}{2} \coth\left({\frac{\tilde{\omega}_{E_m}}{2T_E}}\right) 
    \end{pmatrix},
\end{equation}
As each unit is replaced between collisions, we can neglect the time dependence of $\sigma_E$. The frequencies $\tilde{\omega}_{E_m}$ correspond to the normal mode frequencies of the environmental unit, taking values of 
\begin{equation}
    \tilde{\omega}_{E_m} = \sqrt{\omega_E^2 + 8\lambda_I \sin^2\left[\frac{\pi}{N_E}(m-1)\right]}, \label{eq:Normal Modes}
\end{equation}
with $m=1,2,...,N_E $.  The full derivation of this can be found in \ref{App; N-Oscillator}.
These normal mode frequencies can be understood as the frequencies $\tilde{\omega}_{E_m}$ that satisfy 
\begin{equation}
    H_E = \frac{1}{2}\sum_m (\tilde{\omega}_{E_m}^2 \tilde{x}_{E_m}^2 + \tilde{p}_{E_m}^2).
\end{equation}

Due to the thermal nature of the environmental units and the system's initial state, each element of $\langle R\rangle_{\text{tot}} $ is identically zero and remains zero throughout the propagation as $\langle\dot{R}\rangle_{\text{tot}}$ is described through Eq.~\eqref{eq:R Dynamics}, i.e. as a linear combination of the elements of $\langle R\rangle_{\text{tot}} $. The dynamics of the system can therefore be fully described through the dynamics of its corresponding covariance matrix.

\subsection{Example I: Beamsplitter Interactions} \label{sec; BS Example}
As a first example, we consider the interactions between the system and environment to be beamsplitter-like, such that the interaction Hamiltonian $H_{SE}$ takes the form
\begin{equation}
    \label{eq:HSE}
    H_{SE} = \sum_{i=1}^{N_E} g_{E_i}\left(\sqrt{\omega_S \omega_{E}} x_S x_{E_i} + \frac{p_Sp_{E_i}}{\sqrt{\omega_S\omega_{E}}}\right),
\end{equation}
where $g_{E_i}$ are the coupling constants controlling the strength of the beamsplitter interaction between the system and each environmental oscillator $E_i$. The propagation matrix $D_{\text{tot}}$ can then be described through $D_{\text{tot}} = \Omega_{N_{\rm tot}} M_{\text{tot}}$ \cite{Reid_2017}.

Eq.~\eqref{eq:HSE} can be recast in terms of the normal mode coordinate system $Q_E$, defined in Sec.~\ref{sec; General Structure}, such that 
\begin{equation}
    H_{SE} = \sum_{m=1}^{N_E} \tilde{g}_{E_m}\left(\sqrt{\omega_S \omega_{E}} x_S \tilde{x}_{E_m} + \frac{p_S \tilde{p}_{E_m}}{\sqrt{\omega_S\omega_{E}}}\right).
\end{equation}
Here, we have defined new coupling constants $\tilde{g}_{E_m}$ through
\begin{equation}
    \tilde{g}_{E_m} = \sum_{i=1}^{N_E} G_{mi} g_{E_i}.
\end{equation}
The corresponding Lyapunov equation can then be derived via Eq.~\eqref{eq: Pre continuous}, leading to a differential equation in the continuous time limit, i.e. $\delta t \to 0$,
\begin{equation}
    \dot{\sigma}_S(t) = D_{\rm{BS}} \sigma_S + \sigma_S D_{\rm{BS}}^\intercal + V_{\rm{BS}}, \label{eq: BS ME}
\end{equation}
with 
\begin{equation}
    \label{eq:DBS}
    D_{\rm{BS}} = D_S - \frac{\gamma_{\hat{E}}}{2}\mathds{1}_2
\end{equation}
and 
\begin{equation}
    \label{eq:VBS}
    V_{\rm{BS}} = \sum_{m=1}^{N_E} \frac{\tilde{\gamma}_{E_m}}{2} \coth\left(\frac{\tilde{\omega}_{E_m}}{2T_E}\right)
    \begin{pmatrix}
        \frac{\Tilde{\omega}_{E_m}}{\omega_S \; \omega_E}  & 0 \\
        0 & \frac{\omega_S \; \omega_E}{\Tilde{\omega}_{E_m}}
    \end{pmatrix}.
\end{equation}
Here, $\tilde{\gamma}_{E_m}$ denotes a rescaled coupling constant such that $\tilde{g}_{E_m}^2\delta t \to \tilde{\gamma}_{E_{m}}$ as $\delta t \to 0$, corresponding to the strength of the interactions with an individual, uncoupled normal mode $m$ in the continuous time limit. The effective dissipation rate $\gamma_{\hat{E}}$ is defined through
\begin{equation}
    \gamma_{\hat{E}} = \sum_{m=1}^N \tilde{\gamma}_{E_{m}}. 
\end{equation}

Following the derivations outlined in Sec.~\ref{sec; Thermo Quantities}, we obtain expressions for the transient thermodynamic quantities of interest, namely the flow of internal energy $\dot{U}(t)$, heat $\dot{Q}(t)$ and work $\dot{W}(t)$, described through

\begin{eqnarray}
    \dot{U}(t) &=& \sum_{m=1}^{N_E} \frac{\tilde{\gamma}_{E_m}}{4} \left ( \frac{(\omega_E^2 + \tilde{\omega}_{E_m}^2)\omega_S}{\;\omega_E \;\tilde{\omega}_{E_m}}\coth\left (\frac{\tilde{\omega}_{E_m}}{2T_E}\right ) - 2\omega_S^2 \sigma_{xx}(t) - 2\sigma_{pp}(t)\right ),
    \\
    \dot{Q}(t) &=& \sum_{m=1}^{N_E} \frac{\tilde{\gamma}_{E_m}}{2}  \left ( \tilde{\omega}_{E_m}\coth\left (\frac{\tilde{\omega}_{E_m}}{2T_E}\right ) -\omega_S \; \omega_E \sigma_{xx}(t)- \frac{\tilde{\omega}_{E_m}^2}{\omega_S \; \omega_E}\sigma_{pp}(t)\right ),
    \end{eqnarray}
    and 
    \begin{equation}
        \dot{W}(t) = \dot{U}(t) - \dot{Q}(t).
    \end{equation}

The dynamics of the system are found to be analogous to an effective setup in which the system interacts through a beamsplitter-like interaction via a collision model, with environmental units composed of a single oscillator prepared in a squeezed thermal state with squeezing rate $r$, such that
\begin{equation}
    \langle x_{\hat{E}}^2 \rangle_{\text{sq}} = \frac{1}{2 e^{2r} \omega_E}\coth\left(\frac{\omega_E}{2T_{\hat{E}}}\right)
\end{equation}
and 
\begin{equation}
    \langle p_{\hat{E}}^2 \rangle_{\text{sq}} = \frac{e^{2r}\omega_E}{2}  \coth\left(\frac{\omega_E}{2T_{\hat{E}}}\right),
\end{equation}
where the subscript $\hat{E}$ denotes the single oscillator effective environment and $\langle A \rangle_{\text{sq}}$ denotes the expectation value of an operator $A$ prepared in the squeezed state. 
We find a squeezing parameter of
\begin{equation}
    r = \frac{1}{4}\ln\left( \frac{\alpha_p}{\alpha_x} \right), \label{eq:BS Squeeze}
\end{equation}
where
\begin{equation}
    \alpha_p = \sum_{m=1}^{N_E} \frac{\tilde{\gamma}_{E_m} \tilde{\omega}_{E_m}}{\omega_E} \coth\left(\frac{\tilde{\omega}_{E_m}}{2T_E}\right)
\end{equation}
and 
\begin{equation}
    \alpha_x = \sum_{m=1}^{N_E}  \frac{\tilde{\gamma}_{E_m} \omega_E}{\tilde{\omega}_{E_m}} \coth\left(\frac{\tilde{\omega}_{E_m}}{2T_E}\right).
\end{equation}

As $r>0$ for $\omega_E>\tilde{\omega}_{E_m}$ and $r<0$ for $\omega_E<\tilde{\omega}_{E_m}$ when $m \neq 1$, it is clear that this squeezing decreases the variance of the $x$ quadrature for $\lambda_I < 0$, while $\lambda_I > 0$ decreases the variance of the $p$ quadrature.

The effective temperature $T_{\hat{E}}$ can be shown to be
\begin{equation}
    T_{\hat{E}} = \frac{\omega_E}{2\coth^{-1}\left(\frac{\sqrt{\alpha_x \alpha_p}}{\gamma_{\hat{E}}}\right)}. \label{eq:Eff Temperature}
\end{equation}
This effective temperature $T_{\hat{E}}$ is then controlled by $\lambda_I$, allowing it to be greater than, less than or equal to $T_E$.

These dynamics then follow an effective Lyapunov equation for the system's covariance matrix of
\begin{equation}
    \dot{\sigma}_{S} = D_{\text{eff}}\sigma_S + \sigma_S D_{\text{eff}}^\intercal + V_{\text{eff}}, \label{eq:Master Equation BS}
\end{equation}
where 
\begin{equation}
    D_{\text{eff}} = D_{\rm{BS}},
\end{equation}
and
\begin{equation}
    V_{\text{eff}} = \frac{\gamma_{\hat{E}}}{2}\coth \left(\frac{\omega_E}{2T_{\hat{E}}}\right)
    \begin{pmatrix}
        \frac{e^{2r}}{\omega_S} & 0 \\
        0 & \frac{\omega_S}{e^{2r} } 
    \end{pmatrix}.
\end{equation}

We consider the steady state of the system that satisfies the equation $\dot{\sigma}_S=0$.
The corresponding steady state energy of the system $\langle H_S\rangle_{\text{Steady}}$ reads
\begin{equation}
    \langle H_S\rangle_{\text{Steady}} = \frac{\omega_S}{2} \coth \left(\frac{\omega_E}{2T_{\hat{E}}}\right)\cosh(2r). \label{eq:HSteady}
\end{equation}
As with the effective temperature, the steady state energy is controlled by the choice of $\lambda_I$, and is independent of the initial state of the system.

Although we have assumed a generic model so far, in which the system is potentially coupled to all normal modes of the environmental unit, it is also possible to choose the couplings $g_{E_i}$ in Eq.~\eqref{eq:HSE} such that the system interacts solely with one specific mode with frequency $\tilde\omega_{E_{m'}}$. This is achieved by choosing $g_{E_i} = g G_{m'i} $, where $g$ is the overall coupling strength. With this choice, the system-environment interaction of Eq.~\eqref{eq:HSE} becomes:
\begin{equation}
    H_{SE} =  g \left(\sqrt{\omega_S \omega_{E}} x_S \tilde x_{E_{m'}} + \frac{p_S \tilde p_{E_{m'}}}{\sqrt{\omega_S\omega_{E}}}\right).
\end{equation}

In this case, the expressions for the effective squeezing rate and temperature simplify to
\begin{equation}
    r = \frac{1}{2}\ln\left( \frac{\tilde{\omega}_{E_{m'}}}{\omega_E} \right)
\end{equation}
and
\begin{equation}
    T_{\hat{E}} = \frac{\omega_E}{\tilde{\omega} _{m'}} T_E.
\end{equation}

We shall consider two examples in the following sections: one in which the system interacts exclusively with the center of mass mode frequency $\tilde{\omega}_{\text{CoM}} = \omega_E$, and one in which it interacts with the normal mode frequency corresponding to the largest wavevector within the Brillouin zone and the shortest wavelength supported by the environmental unit.

\subsubsection{Center of Mass}
 The first case of interest is when the system interacts exclusively with the center of mass of each environmental unit, with frequency $\tilde{\omega}_{\rm{CoM}}=\omega_E$, obtained from Eq.~\eqref{eq:Normal Modes} with $m=1$. In this scheme, the system's covariance matrix follows Eq.~\eqref{eq:Master Equation BS}, with a squeezing parameter of $r=0$ (since $\alpha_p=\alpha_x$). The resulting dynamics are equivalent to that of a system interacting with units consisting of a single oscillator of frequency $\omega_E$ prepared in a thermal state at temperature $T_E$ via a collision model through a beamsplitter-like interaction, see for example Ref.~\cite{Leitch_2022}. 
 
 The flow of heat, work and the internal energy are found to be
 \begin{equation}
     \dot{U}(t) = \omega_S f(t),
 \end{equation}
  \begin{equation}
     \dot{Q}(t) = -\omega_E f(t),
 \end{equation}
 and
   \begin{equation}
     \dot{W}(t) = (\omega_S -\omega_E) f(t),
 \end{equation}
 where
 \begin{equation}
    f(t) = \tilde{\gamma}_{E_1} \left[\omega_S \sigma_{xx}(t) + \frac{1}{\omega_S} \sigma_{pp}(t) - \coth\left(\frac{\omega_E}{2T_E} \right) \right ].
 \end{equation}
 
 Immediately, one can see agreement with the first law of thermodynamics (see Eq.~\eqref{eq:firstlaw}). For the case of $\omega_S = \omega_E$, one can recognise an inability to produce or absorb work in the center of mass case,  explained through the commutator $[H_S + H_E, H_{SE}]$. Following Ref.~\cite{De_Chiara_2018}, the condition $[H_S + H_E, H_{SE}] = 0$ is sufficient to ensure $\dot{W}(t)=0$. The fact that  $\dot{U}(t), \dot Q(t)$ and $\dot W(t)$ can be expressed as the product of a common function $f(t)$ and the frequencies $\omega_S$ and $\omega_E$ is reminiscent of similar results for a continuously operated Otto cycle (see for instance Ref.~\cite{Leitch_2022}). This is a consequence of the linearity of the coupling between the harmonic oscillators in the environment and the system. For the $\omega_S = \omega_E$ case, one finds in the steady state
 \begin{eqnarray}
     \dot{U} &=& 0,
     \\
    \dot{Q} &=& 0,
     \\
    \dot{W} &=& 0.
 \end{eqnarray}
 
 Since, in this case, the steady state of the system is a thermal state at temperature $T_E$, it follows that for long times $f(t)\to 0$ because the detailed balance condition is satisfied. Under these conditions $\dot{U}(t), \dot Q(t)$ and $\dot W(t)$ vanish as expected for an equilibrium state.

Due to the lack of squeezing or change in the effective temperature, for $\omega_S = \omega_E$ the dynamics of the system follows the intuitive path: $T_E>T_S$ leads to the system heating until a steady state at temperature $T_E$ is reached; $T_E<T_S$ leads to the system cooling until a steady state at temperature $T_E$ and the $T_E=T_S$ case produces no changes within the system. As there is no contribution from any other normal mode frequencies, and $\tilde{\omega}_{E_{\rm{CoM}}}$ does not depend on $\lambda_I$, it is clear in this case that the structure within the environment does not affect the dynamics or steady state of the system. 

Following Sec.~\ref{sec; Thermo Quantities}, one also finds $\Sigma(t)\geq 0$, showing agreement with the second law of thermodynamics.

\subsubsection{Largest wavevector within the Brillouin zone}
The second case of interest is when the system interacts solely with the normal mode frequency $\tilde{\omega}_{\rm{Max}}$ corresponding to the largest wavevector within the Brillouin zone. This corresponds to the shortest wavelength mode with the frequency $\tilde{\omega}_{E_{m'}}$, where $m'$ is chosen to be the value of $m$ for which the $\sin^2\left[\frac{\pi}{N_E}(m-1)\right]$ term in Eq.~\eqref{eq:Normal Modes} is maximised. This frequency is found to be

\begin{equation}
    \tilde{\omega}_{\text{Max}} = 
    \begin{cases}
        \sqrt{\omega_E^2 + 8\lambda_I},& \text{for even } N_E\\
    \sqrt{\omega_E^2 + 8\lambda_I \cos^2\left(\frac{\pi}{2N_E}\right)},              & \text{for odd } N_E.
    \end{cases} \label{eq: Max Normal Mode}
\end{equation}
 One must take care to choose appropriate values of $\lambda_I$ such that $\tilde{\omega}_{\text{Max}}$ remains real-valued.
 
\begin{figure}[t]
    \centering
    \includegraphics[width=0.7 \linewidth]{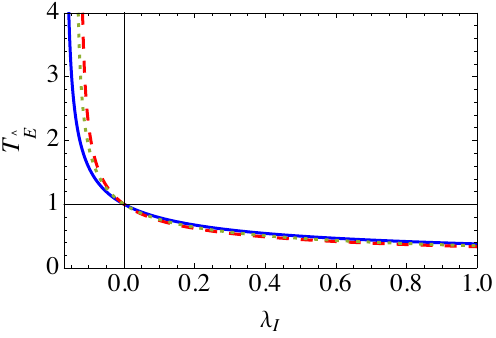}
    \caption{Effective temperature of the environmental units $T_{\hat{E}}$ interacting with the system as a function of the internal coupling strength $\lambda_I$ within the environment. The blue (solid), red (dashed) and green(dotted) lines represent the effective temperature of an environmental unit composed of $3$, $4$ and $5$ oscillators respectively, where the system interacts solely with the normal mode frequency of the environmental units $\tilde{\omega}_{E_{\text{Max}}}$ corresponding to the largest wavevector within the Brillouin zone. Parameters: $\omega_E = 1$, $T_E = 1$, $\tilde \gamma_{E_{m'}} = 0.5$.}
    \label{fig:RM Effective Temperature}
\end{figure}

 In this regime, the system's dynamics can be mapped to that of a quantum harmonic oscillator interacting with environmental units prepared in a squeezed thermal state with effective temperature $T_{\hat{E}}$ (see Eq.~\eqref{eq:Eff Temperature}) and squeezing parameter $r$ (see Eq.~\eqref{eq:BS Squeeze}). This is chosen as a case of particular interest as it allows for control over the effective temperature of the environment. We plot the dependence of the effective temperature $T_{\hat{E}}$  on the environment's internal coupling strength $\lambda_I$ for $N_E=3,4,5$ in Fig.~\ref{fig:RM Effective Temperature}. The temperature $T_{\hat{E}}$ is always monotonically decreasing with $\lambda_I$, such that $T_{\hat{E}}>T_E$ for $\lambda_I < 0$ and $T_{\hat{E}}\le T_E$ for $\lambda_I \le 0$. The dependence of the effective temperature on $N_E$ is very small as $\tilde{\omega}_{\text{Max}}$ depends only weakly on the number of environment oscillators.
In the absence of internal interactions within the environment, i.e. $\lambda_I \to 0$, one recovers the center of mass dynamics with $T_{\hat{E}}=T_E$, analogous to those for beamsplitter-like interactions via a collision model between the system and single oscillator environmental units prepared in a thermal state.

\begin{figure}[ht]
\begin{center}
    \includegraphics[width=0.478 \linewidth]{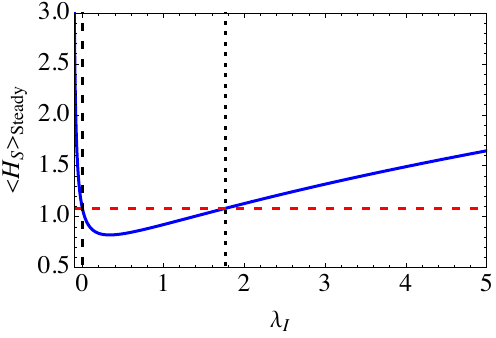}
    \includegraphics[width=0.49 \linewidth]{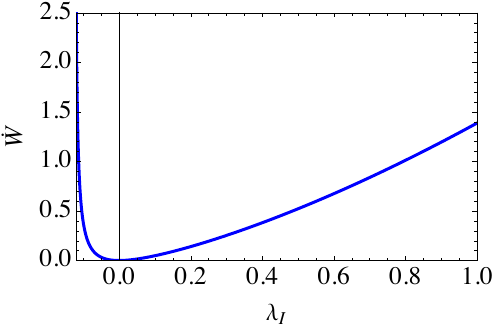}
    \caption{The steady-state energy $H_{\rm{Steady}}$ (left) and work $\dot W$ (right) applied to a system interacting with the normal mode frequency of the environmental units $\tilde{\omega}_{E_{\text{Max}}}$ corresponding to the largest wavevector within the Brillouin zone. Each unit is composed of $4$ oscillators prepared in a thermal state at a temperature $T_E$ as a function of the internal coupling strength $\lambda_I$ (solid line). The horizontal dashed line represents the energy of a thermal state $E_{\rm eq}$. The vertical lines act as a guide to the eye, with the dashed line at $\lambda_I=0$ and the dotted line at $\lambda_{I} = \lambda_{I_C}$. The steady state internal energy is independent of internal coupling, such that $\dot{U}=0$ for all $\lambda_I$, while the heat flow can be found through $\dot{Q} = - \dot{W}$. Parameters: $\omega_S = \omega_E = 1$, $T_E = 1$, $\tilde \gamma_{E_{m'}} = 0.5$.}
    \label{fig:BS Eng}
    \end{center}
\end{figure}

The steady-state energy, $\langle H_S\rangle_{\text{Steady}}$, can be obtained through Eq.~\eqref{eq:HSteady} and compared to the thermal energy $E_{\rm eq}= \omega_S\coth[\omega_E/2T_E]/2$ of a quantum harmonic oscillator in equilibrium with an unstructured environment ($\lambda_I=0$). The plot of Fig.~\ref{fig:BS Eng} shows three different regions of interest. The first of these, in which $\lambda_I<0$, produces a steady-state energy larger than $E_{\rm eq}$, with small changes in $\lambda_I$ leading to large changes in the steady-state energy. The second region, where $0<\lambda_I<\lambda_{I_C}$, produces a steady-state energy lower than the unstructured environment or the center of mass cases. The quantity $\lambda_{I_C}$ is the value $\lambda_I\neq 0$ for which $\langle H_S\rangle_{\text{Steady}}$ is equal to $E_{\rm eq}$. In this region the system is effectively cooled to a temperature lower than that of the environment thanks to the internal structure of the environmental units and their thermal coherences. 
The third region, with $\lambda_I > \lambda_{I_C}$, again produces a steady-state energy larger than the unstructured case, however the magnitude of the change of the steady state energy with respect to changes in $\lambda_I$ is much smaller than that of first region.

Following the derivations outlined in Sec.~\ref{sec; Thermo Quantities}, one can obtain expressions for particular thermodynamic quantities of interest, namely the internal energy $\dot{U}(t)$, heat $\dot{Q}(t)$ and work $\dot{W}(t)$ flows involved in this process. Focusing on the steady state, these quantities can be examined as a function of internal coupling within the environment, $\lambda_I$, with $\dot{U} = 0$, $\dot{Q} = - \dot{W}$ and
\begin{equation}
    \dot{W} =\frac{\gamma_{\hat{E}} 
    \; \omega_S^2(\omega_E^2 - \tilde{\omega}_{E_{\rm{Max}}}^2)^2}{ 2\; \omega_E^2 \; \tilde{\omega}_{E_{\rm{Max}}}(\gamma_E^2 + 4\omega_S^2)} \coth\left(\frac{\tilde{\omega}_{E_{\rm{Max}}}}{2T_E}\right).
\end{equation}
 One finds consistently that $\dot{W}\geq 0$, with the equality holding for $\lambda_I =0$, wherein the dynamics reduce to those of the center of mass regime. Work is therefore always injected into the system for $\lambda_I \neq0$ (see Fig.~\ref{fig:BS Eng}). Power always increases monotonically with $|\lambda_I|$ due to the resulting larger frequency detuning $\omega_E-\tilde{\omega}_{E_{\rm{Max}}}$. Similarly, one finds $\dot{Q}\leq 0$, with the same conditions on the equality, such that for $\lambda_I \neq 0$, heat is always extracted from the system and damped into the environment and the first law of thermodynamics is always satisfied. 
 One can contextualise this into the framework of quantum thermodynamic machine cycles by adding a second environment, interacting with the system through beamsplitter-like interactions via a collision model, with each unit consisting of a single harmonic oscillator. With the inclusion of this secondary environment, one can construct a regime with $\omega_S \neq \omega_E$ in which this model acts as a refrigerator, consuming work for heat to flow from a cold to a hot reservoir. 

Following the procedure outlined in Sec.~\ref{sec; Thermo Quantities}, one finds that while the entropy may increase or decrease to a steady state depending on the choice of $\lambda_I$, the entropy production is always positive, ensuring the second law of thermodynamics is consistently satisfied.

This scheme lends itself well to physical implementation, such as through trapped ion platforms using Coulomb crystals~\cite{MonroeRMP2021}. One can see from Eq.~\eqref{eq: Max Normal Mode} that this model is identical for all even $N_E$, with the exception of the special case $N_E = 2$. Therefore, one only requires $N_E=4$ to achieve the maximum deviation from $\omega_E$ for a given $\lambda_I$. Experimentally, this can be achieved by `resetting' an environmental unit of $4$ oscillators to the initial state between interactions to avoid preparing a large number of units, provided the reset time $\tau_{\text{Reset}}$ is faster than the relaxation time of the system $\tau_{\text{Relax}}$, i.e. $\tau_{\text{Reset}} \ll \tau_{\text{Relax}}$. We emphasise, however, that precise engineering of the environmental units is not crucial to mimic a squeezed bath and that a high-frequency mode coupled to the system's degrees of freedom would be sufficient.

\subsection{Example II: Spring-like Interactions}\label{sec; Spring Example}
The second example presented in this work considers spring-like system-environment interactions, such that $H_{SE}$ takes the form
\begin{equation}
    H_{SE} = \sum_{i=1}^{N_E} \lambda_{E_i}(x_S - x_{E_i})^2,
\end{equation}
where $\lambda_{E_i}$ are the coupling constants controlling the strength of the spring-like interaction between the system and each environmental oscillator $E_i$.

This can be recast in terms of the normal mode coordinate system $Q_E$ defined in Sec.~\ref{sec; General Structure}, as with the beamsplitter example, to produce
\begin{equation}
    H_{SE} = H^{(SE)}_S + H^{(SE)}_E + H^{(SE)}_{SE},
\end{equation}
where $H^{(SE)}_S= \sum_{i=1}^{N_E} \lambda_{E_i} x_S^2$ and gives rise to a renormalised system's frequency $\bar\omega_S^2=\omega_S^2+2\sum_{i=1}^{N_E} \lambda_{E_i}$.
The term $H^{(SE)}_E$ is then given by
\begin{equation}
    H^{(SE)}_E = \sum_{m,m'=1}^{N_E} \tilde \lambda_{E_{mm'}} \tilde{x}_{E_m} \tilde{x}_{E_{m'}}, \label{eq: H_E^SE}
\end{equation}
where $\tilde \lambda_{E_{mm'}}=\sum_{i=1}^{N_E} \lambda_{E_i}G_{mi}  G_{m'i}$ is an effective coupling strength between normal modes, while 
\begin{equation}
    H_{SE}^{(SE)} = - 2\sum_{m=1}^{N_E} \tilde{\lambda}_{E_m} x_S \tilde{x}_{E_m}
\end{equation}
with
\begin{equation}
    \tilde{\lambda}_{E_m} = \sum_{i=1}^{N_E} \lambda_{E_i}G_{mi}.
\end{equation}

The dynamics of the system can then be described in the form of Eq.~\eqref{eq: Pre continuous}. We consider two approaches to study these dynamics. The first is by propagation through discrete time steps wherein $\delta t$ is kept finite. One can begin with

\begin{equation}
    \frac{\sigma_S(t+\delta t) - \sigma_S(t)}{\delta t} 
   = D_{\text{Sp}} \sigma_S(t) + \sigma_S(t) D_{\text{Sp}}^\intercal + \delta t D_S^{\text{RN}}\sigma_S(t) D_S^{{ \intercal \text{RN}}}  + V_{\text{Sp}} +O(\delta t^2), \label{eq:Discrete Lyapunov}
\end{equation}
with
\begin{equation}
    D_\text{Sp} =
    D_S^{\text{RN}} - \frac{\bar{\omega}_S^2 \delta t}{2} \; \mathds{1}_2,
    \label{eq: D_Sp}
\end{equation}
where $D_S^{\text{RN}}$ represents the free evolution of the system's oscillator with the renormalised frequency $\bar\omega_S$, which  depends on the coupling constants $\lambda_{E_i}$. 
The matrix $V_{\rm{Sp}}$ reads 
\begin{equation}
    \label{eq:Vspdiscrete}
    V_{\text{Sp}} =2 \delta t\sum_{m=1}^{N_E}  \tilde{\lambda}_{E_m}^2 
    \begin{pmatrix}
        0 & 0 \\
        0 & \frac{1}{\tilde{\omega}_{E_m}}\coth\left(\frac{\tilde{\omega}_{E_m}}{2T_E}\right)
    \end{pmatrix}.
\end{equation}

Some of the terms in Eq.~\eqref{eq:Discrete Lyapunov} may lead to difficulties in taking the continuous limit, as the method of rescaling used in Sec.~\ref{sec; BS Example} may lead to divergences. In this model, the rescaling method would require taking $\tilde\lambda_{E_m}^2\delta t\to {\rm const.}$ as $\delta t\to 0$. However when doing so, the term $D_S^{\text{RN}}$ would diverge through its dependence on $\bar\omega_S$, due to the divergence of $\tilde\lambda_{E_m}$. We remark, however, that using a finite $\delta t$, as in the discrete evolution, would not lead to any mathematical issue. As the continuous time limit is not possible in the general case, we instead propagate the system through Eq.~\eqref{eq:sigma(t + dt)} to avoid the approximation due to the Taylor expansion in $\delta t$.

 While the discrete propagation model has been presented as generic so far, we shall focus on interactions with the center of mass frequency, as this case requires discrete propagation. Such interactions, i.e. those with an individual chosen normal mode frequency $\tilde\omega_{E_{m'}}$, can be achieved by choosing couplings $\lambda_{E_i} = \lambda_E G_{m'i} $, where $\lambda_E$ is the overall coupling strength.
 
For the second approach, we will consider the continuous limit $\delta t\to 0$ in the specific case $\sum_i^{N_E}\lambda_{E_i} = 0$ which leaves the system's frequency unaltered, i.e. $\bar\omega_S=\omega_S$, avoiding any divergence. 

\subsubsection{Discrete Time}
As discussed above, for generic $\lambda_{E_i}$ in the spring-like interaction model, it is not possible to derive a continuous Lyapunov equation for the covariance matrix akin to Eq.~\eqref{eq:Master Equation BS}.
However, one can still propagate the system through discrete time steps $\delta t$ according to Eq.~\eqref{eq:sigma(t + dt)}. Here, we present this discrete-time evolution resulting from interactions between a single oscillator system, prepared initially in a thermal state at temperature $T_S$ and frequency $\omega_S$, and the center of mass mode of the environmental units, each composed of $N_E = 4$ oscillators, i.e. we choose homogeneous interaction couplings $\lambda_{E_i}= \lambda_E G_{m'i}$ with $m'=1$. Furthermore we assume each environmental unit to be initially prepared in a thermal state at a temperature $T_E$.

\begin{figure}[ht]
\begin{center}
    {\includegraphics[width=0.7\linewidth]{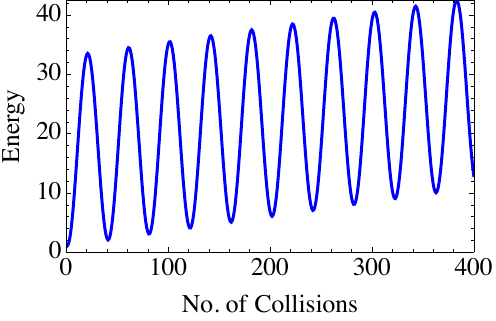}}
    \caption{Dynamics of the energy of a single oscillator system prepared in the thermal state as a function of time, with discrete time-steps $\delta t = 0.01$, interacting with the center of mass frequency of environmental units composed of $N_E=4$ oscillators. Parameters: $\omega_S=\omega_E = 1$, $T_S=T_E=1$, $\lambda_I=0.67$, $\tilde \lambda_{{E_{1}}} = 15$.}
    \label{fig:Discrete Spring Dynamics}
    \end{center}
\end{figure}

We propagate the system's covariance matrix numerically using Eq.~\eqref{eq:sigma(t + dt)}, with the corresponding results for the system's energy shown in Fig.~\ref{fig:Discrete Spring Dynamics}. We observe that, unlike the beamsplitter case, the system does not tend to a steady state. Instead, the energy oscillates around a linearly increasing average. This originates from a persistent injection of energy from the environment to the system due to the spring-like form of interaction.

\begin{figure}[ht]
\begin{center}
    \includegraphics[width=0.49 \linewidth]{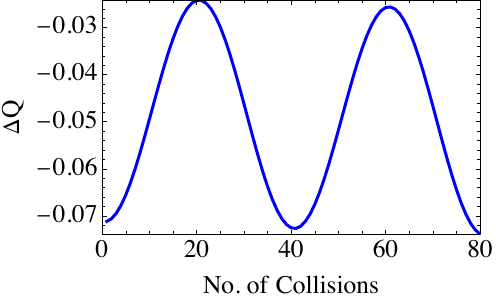}
    \includegraphics[width=0.46 \linewidth]{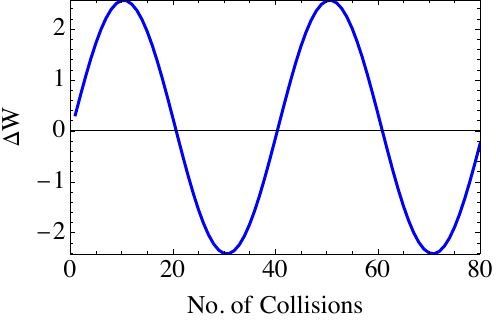}
    \caption{The heat current $\Delta Q$ (left) and work $\Delta W$ (right) applied to the system the system with same parameters as Fig.~\ref{fig:Discrete Spring Dynamics}.
    The change in the internal energy of the system $\Delta U$ can be found as the sum $\Delta Q$ and $\Delta W$.}
    \label{fig:Spring Discrete Thermo QW}
    \end{center}
\end{figure}

This is confirmed by looking at the heat, $\Delta Q(t)$, and work, $\Delta W(t)$, per collision computed using the methods outlined in Sec.~\ref{sec; Thermo Quantities}. As shown in Fig.~\ref{fig:Spring Discrete Thermo QW}, work is periodically injected and extracted from the system, with a net positive injection over long time frames, accounting for the increase in the systems energy between each period. One also finds  dissipation through the heat $\Delta Q$, which remains negative throughout the propagation.

The entropy of the system $S(t)$ also increases as one would expect, as the system's energy increases throughout the process. Once again, the second law of thermodynamics can be verified by examining the entropy production $\Sigma(t)$, where one requires $\Sigma(t)\geq 0$.

\subsubsection{Continuous Time}

It is possible to choose interaction constants $\lambda_{E_i}$ such that divergences do not appear, allowing for the continuous-time limit to be taken~\cite{PhysRevA.97.052120}. Applying the rescaling $\lambda_{E_i}\lambda_{E_j}\delta t \to \Lambda_{E_{ij}}$ as $\delta t \to 0$, can in general produce $\mathcal{L}_1(\sigma_{\text{tot}})\to \infty$  however this can be avoided provided $\sum_{i=1}^{N_E}\lambda_{E_i} = 0$.

The resulting Lyapunov equation can then be stated  as 
\begin{equation}
    \dot{\sigma}_{S} = D_S\sigma_S + \sigma_S D_S^\intercal + V_{\text{Sp}}, \label{eq:Master Equation Sp}
\end{equation}
where $V_{\text{Sp}}$ is given by Eq.~\eqref{eq:Vspdiscrete} after replacing $\tilde{\lambda}_{E_m}^2 \delta t \to \tilde{\Lambda}_{E_{m}}$ and taking the continuous limit $\delta t \to 0$. Notice that the trick we employed to obtain an effective continuous time dynamics would not be possible if the environmental units consisted of only a single oscillator through solely spring-like interactions outside of the trivial $\lambda_{E_i} = 0$ case, as a minimum of two oscillators are necessary to fulfill the condition of $\sum_{i=1}^{N_E}\lambda_{E_i} = 0$.

\begin{figure}[ht]
\begin{center}
    \includegraphics[width=0.49 \linewidth]{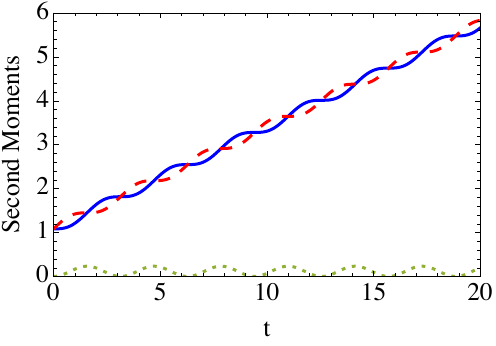}
    \includegraphics[width=0.49 \linewidth]{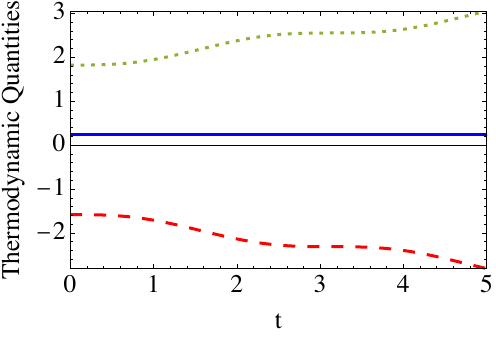}
    \caption{Dynamics of a system interacting with the normal mode frequency $\tilde{\omega}_{E_{\text{Max}}}$ corresponding to the largest wavevector within the Brillouin zone in the continuous-time limit. The second moments (left) and thermodynamic quantities (right) are plotted as a function of time $t$. The second moments are described by the elements of the covariance matrix, particularly $\langle x_S^2 \rangle$ (blue, solid),  $\langle p_S^2 \rangle$ (red, dashed)) and $\langle x_S p_S + p_S x_S \rangle$ (green, dotted). The thermodynamic quantities presented are the flow of internal energy $\dot U(t)$ (blue, solid), heat $\dot Q(t)$ (red, dashed) and work $\dot W(t)$ (green, dotted). Parameters: $\omega_S = \omega_E = 1$, $T_S = T_E = 1$, $\lambda_I = 0.67$, $\tilde{\Lambda}_{E_{m'}} = 0.5$.}
    \label{fig:Spring RM}
    \end{center}
\end{figure}

To maximise the effects of the environment, we present the dynamics due to interactions of the system with the normal mode frequency $\tilde{\omega}_{E_{\text{Max}}}$ described in Eq.~\eqref{eq: Max Normal Mode}.
In Fig.~\ref{fig:Spring RM},  we plot the second moments, $\langle x^2_S \rangle$ and $\langle p_S^2 \rangle$, which oscillate around a linearly increasing average value. The almost linear increase in these quantities is reminiscent of a diffusion model, in which the variances increase linearly with respect to time. 
Similarly to the discrete case, this is due to a persistent energy injection from the environment.

One can again calculate the thermodynamic quantities described in Sec.~\ref{sec; Thermo Quantities} to find that work must always be injected into the system, as $\dot{W}(t) > 0$, while heat consistently flows out of the system, as $\dot{Q}(t) < 0$, also shown in Fig.~\ref{fig:Spring RM}. The internal energy increases linearly over time, with $\dot{U}(t)$ equal to a positive constant (whose expression can be found but it does not give additional information). Because of this linear increase in the internal energy, it is clear that a model using solely spring-like interactions does not admit a steady state. 

One may investigate the entropy and entropy production of the system in this case, resulting again in the verification of the second law of thermodynamics, wherein the change in both the entropy $S(t)$ and entropy production $\Sigma(t)$ between collisions is positive.

It is possible to induce a steady state through the use of a secondary environment. As a simple example, we assume this additional environment to consist of units composed of a single oscillator in a thermal state, with frequency $\omega_B$ and temperature $T_B$ interacting with the system through beamsplitter-like interactions via a collision model, as in Sec.~\ref{sec; BS Example} ($N_E=1$ and $\lambda_I=0$), see Eqs.\eqref{eq: BS ME},\eqref{eq:DBS},\eqref{eq:VBS}. 

The total propagation takes the form of
\begin{equation}
    \dot{\sigma}_S(t) = D_L \sigma_S(t) + \sigma_S(t) D^\intercal_L + V_{\text{Sp}} + V_B, \label{eq:Lyapunov EoM}
\end{equation}
with
\begin{equation}
    D_L = D_S - \frac{\gamma_B}{2}\mathds{1}_2,
\end{equation}
and
\begin{equation}
    V_B = 
    \frac{1}{2}
    \begin{pmatrix}
        \frac{\gamma_B}{\omega_S}\coth\left(\frac{\omega_B}{2T_B}\right) & 0 \\
        0 & \gamma_B\omega_S \coth\left(\frac{\omega_B}{2T_B}\right)
    \end{pmatrix},
\end{equation}
where $\gamma_B$ is related to the coupling of this secondary environment to the system.    

\begin{figure}[ht]
\begin{center}
    \includegraphics[width=0.478\linewidth]{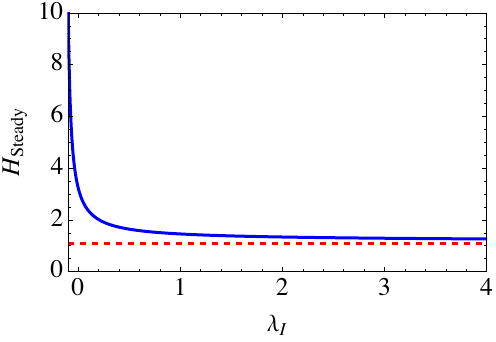}
    \includegraphics[width=0.49 \linewidth]{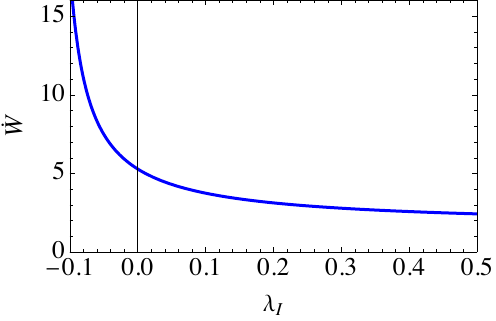}
    \caption{The steady state energy $H_{\rm{Steady}}$ (left) and work $\dot W$ (right) as a function of the internal environmental coupling strength $\lambda_I$ of a structured primary environment. The system also interacts with a secondary 'drain' environment. Each unit of the primary environment is composed of $N_E=4$ oscillators interacting with the system via spring-like interactions. The units of the secondary environment are composed of $N_E = 1$ oscillator interacting with the system via beamsplitter-like interactions. In the steady-state, $\dot{U}=0$ for all $\lambda_I$, and the heat flow can be found through $\dot{Q} = - \dot{W}$. The initial energy (dashed, red) of the system with these parameters is also plotted. Parameters: $\omega_S = \omega_E = \omega_B = 1$, $T_S = T_E = T_B = 1$, $\tilde{\Lambda}_{E_{m'}} = 0.5$, $\gamma_B =0.5$.
    \label{fig:Spring Drain}}
\end{center}
\end{figure}

In this case, the system tends to a steady state described through
\begin{equation}
    \langle H_S\rangle_{\text{Steady}} = \sum_{m=2}^{N_E}\frac{\tilde{\Lambda}_{E_m}}{\gamma_B \tilde{\omega}_{E_m}}\coth\left(\frac{\tilde{\omega}_{E_m}}{2T_E}\right) +
    \frac{\omega_S}{2}\coth\left(\frac{\omega_B}{2T_B}\right). \label{eq:Drain Steady State}
\end{equation}
From Fig.~\ref{fig:Spring Drain}, one can see that the primary environment interacting with the system via spring-like interactions injects a constant amount of energy into the system with each collision. For the secondary environment interacting via beamsplitter-like interactions, the amount of energy injected or extracted by each unit is dependent on the energy of the system at the time of collision. The steady state then occurs when the rate at which energy is drained by the secondary environment is equal to the constant amount of energy injected by the primary environment. When both the primary and secondary environment are initially at the same temperature with oscillators of equal frequency, taking $\lambda_I = 0$ still results in an increase to the system's steady-state energy. The coupling $\lambda_I$ therefore does not initiate the change, but rather allows for control of its magnitude, where $\lambda_I < 0$ allows for an increase in the steady-state energy that is large in magnitude, while $\lambda_I>0$ allows for an increase that is small and controlled. 

In the steady state, one finds a dissipator-like behaviour, where work is constantly injected into the system, as $\dot W >0$, while heat flows out of the system, as $\dot Q < 0$. In such a steady state, one finds the internal energy is constant, i.e. $\dot U = 0$.

\section{Conclusions And Outlook} \label{sec; Conclusions}
This work presents a general method of designing a collision model that includes structure within the environmental units using continuous-variable quantum systems, namely harmonic oscillators. In some cases, it is possible to construct a Lyapunov equation for the continuous-time evolution of the system's covariance matrix. 
We demonstrated the effect of an internal structure within the environment may have over the system's dynamics and thermodynamics, particularly with respect to the effective temperature of the environmental units as a function of their internal coupling strength $\lambda_I$. 
The dynamics of a system interacting through beamsplitter-like interactions via a collision model with units composed of spring-like coupled oscillators, initially prepared in a thermal state, were found to be analogous to the dynamics due to interactions with single oscillator environmental units prepared in a squeezed thermal state. The magnitude of this squeezing also corresponded to a function of the internal coupling strength. This effective squeezing arises naturally from environmental units prepared in thermal equilibrium, thus considered free within the resource theory of quantum thermodynamics. This is in stark contrast to previous works which considered environmental units which are artificially prepared in squeezed states~\cite{PhysRevA.110.052421,PhysRevA.97.052120,correa2014quantum}, thus necessarily employing additional thermodynamic resources.  

In other cases, in which the system interacts with the environment through spring-like interactions, we found that discrete time propagation is necessary, as taking the continuous-time limit leads to divergences. In this example, the system does not inherently reach a steady state due to constant injections of energy from the environment with each collision.
Nevertheless, we showed that one can include a secondary thermal environment (corresponding to a collision model whose units are made of a single oscillator and interacting with the system via beamsplitter-like interactions) to lead the system to a steady state. 

In all cases, we proved that the first and second laws of thermodynamics are fulfilled. This shows once more that structured collision models can play an important role in the context of thermodynamic investigations, as they allow for a thermodynamically consistent description even in presence of coherence, in contrast with other models where thermodynamic consistence must be enforced.

Verification of these results may be currently achievable through various experimental methods, such as through trapped ions~\cite{Hou2024, Home_2011, Brown2011} or photonic systems~\cite{cuevas_all-optical_2019,PhysRevA.91.012122}. The effect of the topology of the structure within the environmental units would be of keen interest for further study, alongside the effects of varying the initial states, interactions between the system and environment, and of course the internal interactions within the environmental units. Finally, it would be of practical interest to examine the effects of a structured environment to the performances of heat engines and refrigerators, as well as in the charging process of quantum batteries.

A. C. acknowledges support from the UK EPSRC through Grant No. EP/W52444X/1.
S. C. acknowledges support from the PNRR MUR project PE0000023NQSTI.
G.D.C. acknowledges support from the UK EPSRC through Grant No. EP/S02994X/1.

\appendix
\section*{Appendices}

\section{Superoperator Derivation}\label{App; System superoperators}
The derivation of the superoperators relating the total dynamics of the system and environment to the reduced dynamics of solely the system, $\mathcal{L}_0(\sigma_{\text{tot}}(t))$, $\mathcal{L}_1(\sigma_{\text{tot}}(t))$, $\mathcal{L}_2(\sigma_{\text{tot}}(t))$ can be found by considering the set of elements $i$, $j$ of the matrices produced by the total dynamic operators $\mathcal{U}_0(\sigma_{\text{tot}}(t))$, $\mathcal U_1(\sigma_{\text{tot}}(t))$, $\mathcal U_2(\sigma_{\text{tot}}(t))$, for which $i=1,2...,  2N_S$ and $j=1,2..., 2N_S$. More intuitively, one can consider this as selecting only the ``top-left" $2N_S \times 2N_S$ block of the total system-environment covariance matrix after applying the total evolution operators. The action of extracting the properties and dynamics of solely the system will be denoted here as $[\cdot]_S$, such that $\mathcal{L}_i(\sigma_{\text{tot}}(t)) = [\mathcal U_i(\sigma_{\text{tot}}(t))]_S$.

To find $\mathcal{L}_0(\sigma_{\text{tot}}(t))$, one can recognise that the corresponding operator $\mathcal U_0(\sigma_{\text{tot}}(t))$ can be described simply as the application of the identity $\mathds{1}_{2N_{\text{tot}}}$ to $\sigma_{\text{tot}}(t)$, hence,
\begin{equation}
\mathcal{L}_0(\sigma_S (t)) = \sigma_S(t).
\end{equation}

The behaviour due to $\mathcal{L}_1(\sigma_{S}(t))$ can be found by separating the propagation matrix $D_{\text{tot}}$ into interactions due the the system, $D_S'=D_S\oplus O_{2N_E}$, the environment, $D_E'=O_{2N_S} \oplus D_E$ and the system-environment interactions, $D_{SE}$, each of dimension $2N_{\rm{tot}}$, such that 
\begin{equation}
     D_{\rm{tot}} = D_S' + D_E' + D_{SE}, \label{eq: D Expanded}  
\end{equation}
with $O_N$ denoting the zero matrix of dimension $N$.

In order to describe the system's dynamics due to $\mathcal U_1(\sigma_{\text{tot}}(t))$, we consider the contributions from each of these terms separately. First, we see the $D_E$ terms do not contribute to the system's dynamics by recognising $D_{E_{ij}}' = 0$  for all $i=1,2,...,2N_S$, and for $j=1,2,...,N_S$, and therefore 
\begin{eqnarray}
    \sum_k D'_{E_{ik}} \sigma_{\text{tot}_{kj}}(t) &=& 0; \qquad 1 \leq i \leq 2N_S,
    \\ 
    \sum_k \sigma_{\text{tot}_{ik}}(t) {D'}^\intercal_{E_{kj}} &=& 0; \qquad 1 \leq j \leq 2N_S.
\end{eqnarray}

 One can treat the $D_{SE}$ terms using the notation of $[\cdot]_S$, retaining only the $[D_{SE} \sigma_{\text{tot}}(t)]_S$ and $[\sigma_{\text{tot}}(t)D_{SE}^\intercal]_S$ contributions. Similarly, applying this operation on the $D_S'$ terms produces $[D'_{S} \sigma_{S}(t)]_S = D_S \sigma_S(t)$ and $[\sigma_{\text{tot}}(t)D_S'^\intercal]_S = \sigma_S(t) D_S^\intercal$. The dynamics of the system due to the first order $\delta t$ terms can then be written as

\begin{equation}
    \mathcal{L}_1(\sigma_{S}(t)) = \delta t\left \{D_S \sigma_S(t) + \sigma_S(t)D^\intercal_S + [D_{SE}\sigma_{\text{tot}}(t)]_S + [\sigma_{\text{tot}}(t)D_{SE}^\intercal]_S\right\}.
\end{equation}

The second-order terms can be treated through the expansion of $D_{\rm{tot}}$ described in Eq.~\eqref{eq: D Expanded}. The terms involving solely the environmental dynamics $D_E$ and the system once again provide no contribution, as
\begin{eqnarray}
        \sum_{kl} D'_{E_{ik}}D'_{E_{kl}} \sigma_{\text{tot}_{lj}}(t) = 0; \qquad 1 \leq i \leq 2N_S, 
        \\
        \sum_{kl}\sigma_{\text{tot}_{ik}}(t){D'}^\intercal_{E_{kl}}{D'}^\intercal_{E_{lj}} = 0; \qquad 1 \leq j \leq 2N_S,
        \\
        \sum_{kl} D'_{E_{ik}}\sigma_{\text{tot}_{kl}}(t){D'}^\intercal_{E_{lj}} = 0; \qquad 1 \leq i \leq 2N_S.
\end{eqnarray}
One also finds no contribution from terms consisting solely of $D_S$ and $D_E$ terms, as $D'_S D'_E = D'_E D'_S = O_{2N_{\rm{tot}}}$, and
\begin{eqnarray}
    \sum_{kl} D'_{S_{ik}}\sigma_{\text{tot}_{kl}}(t) {D'}^\intercal_{E_{lj}} &=& 0; \qquad 1 \leq i \leq 2N_S,
\\
    \sum_{kl} D'_{E_{ik}}\sigma_{\text{tot}_{kl}}(t) {D'}^\intercal_{S_{lj}} &=& 0; \qquad 1 \leq j \leq 2N_S.
\end{eqnarray}

Finally, there are no contributions from the terms containing solely $D_{SE}$ and $D'_E$, as
\begin{eqnarray}
    \sum_{kl} {D_{E_{ik}}'}{D_{SE_{kl}}}\sigma_{\text{tot}_{lj}}(t)   = 0; \qquad 1 \leq i \leq 2N_S,
\\
    \sum_{kl} \sigma_{\text{tot}_{ik}}(t){D'}^\intercal_{E_{kl}}{D}^\intercal_{SE_{lj}}  = 0; \qquad 1 \leq i \leq 2N_S,
\\
    \sum_{kl} D'_{E_{ik}}\sigma_{\text{tot}_{kl}}(t) {D}^\intercal_{SE_{lj}} = 0; \qquad 1 \leq i \leq 2N_S,
\\
    \sum_{kl} {D_{SE_{ik}}}{D_{E_{kl}}'}\sigma_{\text{tot}_{lj}}(t) = 0; \qquad 1 \leq j \leq 2N_S,
\\
    \sum_{kl} \sigma_{\text{tot}_{ik}}(t) {D}^\intercal_{SE_{kl}}{D'}^\intercal_{E_{lj}} = 0; \qquad 1 \leq j \leq 2N_S,
\\
    \sum_{kl} D_{SE_{ik}}\sigma_{\text{tot}_{kl}}(t) {D'}^\intercal_{E_{lj}}= 0; \qquad 1 \leq j \leq 2N_S.
\end{eqnarray}

The remaining nonzero terms are those related to the unitary dynamics of the system, terms involving the system dynamics and the system-environment interaction, and terms solely involving the system-environment interaction. The second order terms of the system's dynamics with respect to $\delta t$ are then given by 

\begin{equation}
    \mathcal{L}_2(\sigma_S (t)) = \frac{\delta t^2}{2} \left(\mathcal{L}^S_2(\sigma_S (t)) + \mathcal{L}^{SE}_2(\sigma_S (t))+ \mathcal{L}^{S/SE}_2(\sigma_S (t))\right),
\end{equation}
with 
\begin{equation}
    \mathcal{L}^S_2(\sigma_S (t)) = 2D_S \sigma_S(t) D_S^\intercal + D_S^2 \sigma_S(t) + \sigma_S(t)D_S^{\intercal 2},
\end{equation}

\begin{equation}
    \mathcal{L}^{SE}_2(\sigma_S (t)) = 2[D_{SE}\sigma_{\text{tot}}(t)D^\intercal_{SE}]_{S} + [{D}^2_{SE}\sigma_{\text{tot}}(t)]_{S} + [\sigma_{\text{tot}}(t) {D^\intercal}_{SE}^2]_{S},
\end{equation}
and 
\begin{eqnarray}
    \mathcal{L}^{S/SE}_2(\sigma_S (t)) &=& 2[D'_{S}\sigma_{\text{tot}}(t){D^\intercal}_{SE}]_{S} + 2[D_{SE}\sigma_{\text{tot}}(t){D'}^\intercal_{S}]_{S} + [{D_S' D_{SE}} \sigma_{\text{tot}}(t)]_{S} \notag \\ &+& [{D_{SE} D_{S}'} \sigma_{\text{tot}}(t)]_{S} + [\sigma_{\text{tot}}(t){D'}^\intercal_{S}{D}^\intercal_{SE}]_{S} + [\sigma_{\text{tot}}(t){D}^\intercal_{SE}{D'}^\intercal_{S}]_{S}
\end{eqnarray}

\section{Environmental Unit Normal-Mode Decomposition} \label{App; N-Oscillator}
Decomposing the structured environmental units into their corresponding normal modes  greatly simplifies the derivation of the system's dynamics. Here, we consider an environmental unit of $N_E$ oscillators arranged in a ring, coupled via spring-like interactions, with a total Hamiltonian $H_E$ described by Eq.~\eqref{eq:Environment Hamiltonian}, \eqref{eq:H_E Free} and \eqref{eq:H_EI}, such that
\begin{equation}
    H_E = \sum_{n=1}^{N_E} \left [\frac{1}{2}(\omega_E^2x_{E_n}^2 + p_{E_n}^2) +\lambda_I(x_{E_n} - x_{E_{n+1}})^2\right ],
\end{equation}
with periodic boundary conditions ($N_E+1\equiv 1$).
Furthermore, since the kinetic-energy terms are already non-interacting (and the environmental oscillators have all the same mass), we can look for a transformation to a basis in which the position operators are non-interacting, i.e. focusing only on the terms of the Hamiltonian corresponding to the potential energy, 
$$
H_{E_x}=\sum_{n=1}^{N_E} \left [\frac{1}{2}\omega_E^2x_{E_n}^2  +\lambda_I(x_{E_n} - x_{E_{n+1}})^2\right ].
$$
This simplifies the search for a description of the normal modes, leaving only the constraints that the transformation must be symplectic, without inducing interactions between the momentum operators. 

We can obtain the normal modes via a Fourier decomposition described by
\begin{equation}
    x_j = \frac{1}{\sqrt{N_E}}\sum_{m=0}^{N_E - 1}e^{-ijk_m}\tilde{x}_m, \label{eq: Normal Coordinates}
\end{equation}
where $k_m=\frac{2\pi}{N_E}m$ for $m = 0,1, 2, \dots,N_E - 1$ and $\tilde{x}_m$ denotes the normal mode coordinates. As all operators in this appendix refer to the environment, the subscripts $E$ have been dropped for brevity.

The potential-energy Hamiltonian $H_{E_x}$ can then be rewritten in terms of the normal mode coordinates through Eq.~\eqref{eq: Normal Coordinates}, producing
\begin{equation}
    H_{E_x} = \frac{1}{N_E}\sum_{j, m, m'}e^{-ij(k_m+k_{m'})}\left [\frac{\omega_E^2}{2}\tilde{x}_{m}\tilde{x}_{m'} + \lambda_I (1 + e^{-i(k_m + k_{m'})} - e^{-ik_m} - e^{-ik_{m'}})\tilde{x}_m\tilde{x}_{m'}  \right ].
\end{equation}
This can be simplified further through the standard result
\begin{equation}
    \sum_{j=1}^{N_E} e^{-ij(k_m+k_{m'})} = N_E\delta_{k_m,-k_{m'}},
\end{equation}
to obtain
\begin{equation}
    H_{E_x} = \sum_{m=0}^{N_E - 1}\left (\frac{\omega_E^2}{2} +2\lambda_I - \lambda_I(e^{ik_m} + e^{-ik_m})\right )\tilde{x}_m\tilde{x}_{-m} .
\end{equation}
Finally, using trigonometric relations and noting both $\tilde{x}_{-m} = \tilde{x}_m ^\dagger$ and $\tilde{x}_m^\dagger = \tilde{x}_m$, $H_{E_x}$ can be rewritten as
\begin{equation}
    H_{E_x} = \sum_{m=1}^{N_E}\left [\frac{\omega_E^2}{2} + 4\lambda_I \sin^2\left (\frac{\pi}{N_E}(m-1)\right )\right ]\tilde{x}_m^2.
\end{equation}
 One can compare this to the potential-energy Hamiltonian of a set of uncoupled oscillators of frequency $\tilde{\omega}_{E_m}$ to find the frequency of the normal modes, i.e.
\begin{equation}
    \frac{\tilde{\omega}^2_{E_m}}{2} = \frac{\omega_E^2}{2} + 4\lambda_I \sin^2\left (\frac{\pi}{N_E}(m-1)\right ), \label{eq:NM Freqs}
\end{equation}
such that Eq.~\eqref{eq:Normal Modes} is satisfied. 

Eq.~\eqref{eq:NM Freqs} is valid for all $N_E>2$, however it is important to note the special case of $N_E = 2$. Typically, one does not impose ring-like periodic boundary conditions in this case, but rather the oscillators are arranged in a line structure. There are therefore $N_E-1$ interactions in this case (just 1), rather than $N_E$, resulting in a center of mass frequency $\tilde{\omega}_{\text{CoM}} = \omega_E$ and a relative motion frequency of $\tilde{\omega}_{\text{RM}} = \sqrt{4\lambda_I + \omega_E^2}$.

\section*{References}
\bibliographystyle{iopart-num}
\bibliography{bibliography}{}

\end{document}